\documentclass[
    ,final            % use final for the camera ready runs
  ,numberedheadings % uncomment this option for numbered sections
  ]
  {aipproc}

\usepackage{natbib}

\layoutstyle{8x11single}

\begin{document}

\newcommand\aj{{AJ}}%
          % Astronomical Journal
\newcommand\actaa{{Acta Astron.}}%
  % Acta Astronomica
\newcommand\araa{{ARA\&A}}%
          % Annual Review of Astron and Astrophys
\newcommand\apj{{ApJ}}%
          % Astrophysical Journal
\newcommand\apjl{{ApJ}}%
          % Astrophysical Journal, Letters
\newcommand\apjs{{ApJS}}%
          % Astrophysical Journal, Supplement
\newcommand\ao{{Appl.~Opt.}}%
          % Applied Optics
\newcommand\apss{{Ap\&SS}}%
          % Astrophysics and Space Science
\newcommand\aap{{A\&A}}%
          % Astronomy and Astrophysics
\newcommand\aapr{{A\&A~Rev.}}%
          % Astronomy and Astrophysics Reviews
\newcommand\aaps{{A\&AS}}%
          % Astronomy and Astrophysics, Supplement
\newcommand\azh{{AZh}}%
          % Astronomicheskii Zhurnal
\newcommand\baas{{BAAS}}%
          % Bulletin of the AAS
\newcommand\caa{{Chinese Astron. Astrophys.}}%
  % Chinese Astronomy and Astrophysics
\newcommand\cjaa{{Chinese J. Astron. Astrophys.}}%
  % Chinese Journal of Astronomy and Astrophysics
\newcommand\icarus{{Icarus}}%
  % Icarus
\newcommand\jcap{{J. Cosmology Astropart. Phys.}}%
  % Journal of Cosmology and Astroparticle Physics
\newcommand\jrasc{{JRASC}}%
          % Journal of the RAS of Canada
\newcommand\memras{{MmRAS}}%
          % Memoirs of the RAS
\newcommand\mnras{{MNRAS}}%
          % Monthly Notices of the RAS
\newcommand\na{{New A}}%
  % New Astronomy
\newcommand\nar{{New A Rev.}}%
  % New Astronomy Review
\newcommand\pra{{Phys.~Rev.~A}}%
          % Physical Review A: General Physics
\newcommand\prb{{Phys.~Rev.~B}}%
          % Physical Review B: Solid State
\newcommand\prc{{Phys.~Rev.~C}}%
          % Physical Review C
\newcommand\prd{{Phys.~Rev.~D}}%
          % Physical Review D
\newcommand\pre{{Phys.~Rev.~E}}%
          % Physical Review E
\newcommand\prl{{Phys.~Rev.~Lett.}}%
          % Physical Review Letters
\newcommand\pasa{{PASA}}%
  % Publications of the Astron. Soc. of Australia
\newcommand\pasp{{PASP}}%
          % Publications of the ASP
\newcommand\pasj{{PASJ}}%
          % Publications of the ASJ
\newcommand\qjras{{QJRAS}}%
          % Quarterly Journal of the RAS
\newcommand\rmxaa{{Rev. Mexicana Astron. Astrofis.}}%
  % Revista Mexicana de Astronomia y Astrofisica
\newcommand\skytel{{S\&T}}%
          % Sky and Telescope
\newcommand\solphys{{Sol.~Phys.}}%
          % Solar Physics
\newcommand\sovast{{Soviet~Ast.}}%
          % Soviet Astronomy
\newcommand\ssr{{Space~Sci.~Rev.}}%
          % Space Science Reviews
\newcommand\zap{{ZAp}}%
          % Zeitschrift fuer Astrophysik
\newcommand\nat{{Nature}}%
          % Nature
\newcommand\iaucirc{{IAU~Circ.}}%
          % IAU Cirulars
\newcommand\aplett{{Astrophys.~Lett.}}%
          % Astrophysics Letters and Communications
\newcommand\apspr{{Astrophys.~Space~Phys.~Res.}}%
          % Astrophysics Space Physics Research
\newcommand\bain{{Bull.~Astron.~Inst.~Netherlands}}%
          % Bulletin Astronomical Institute of the Netherlands
\newcommand\fcp{{Fund.~Cosmic~Phys.}}%
          % Fundamental Cosmic Physics
\newcommand\gca{{Geochim.~Cosmochim.~Acta}}%
          % Geochimica Cosmochimica Acta
\newcommand\grl{{Geophys.~Res.~Lett.}}%
          % Geophysics Research Letters
\newcommand\jcp{{J.~Chem.~Phys.}}%
          % Journal of Chemical Physics
\newcommand\jgr{{J.~Geophys.~Res.}}%
          % Journal of Geophysical Research
\newcommand\jqsrt{{J.~Quant.~Spec.~Radiat.~Transf.}}%
          % Journal of Quantitiative Spectroscopy and Radiative Trasfer
\newcommand\memsai{{Mem.~Soc.~Astron.~Italiana}}%
          % Mem. Societa Astronomica Italiana
\newcommand\nphysa{{Nucl.~Phys.~A}}%
          % Nuclear Physics A
\newcommand\physrep{{Phys.~Rep.}}%
          % Physics Reports
\newcommand\physscr{{Phys.~Scr}}%
          % Physica Scripta
\newcommand\planss{{Planet.~Space~Sci.}}%
          % Planetary Space Science
\newcommand\procspie{{Proc.~SPIE}}%
          % Proceedings of the SPIE

\title{The effects of surface roughness on lunar Askaryan pulses}


\author{C.W.~James}{
  address={ECAP, Friedrich-Alexander University of Erlangen-Nuremberg, Erwin-Rommel-Str.~1, 91058 Erlangen, Germany},
  email={clancy.james@physik.uni-erlangen.de},}

\begin{abstract}
The effects of lunar surface roughness, on both small and large scales, on Askaryan radio pulses generated by particle cascades beneath the lunar surface has never been fully estimated. Surface roughness affects the chances of a pulse escaping the lunar surface, its coherency, and the characteristic detection geometry. It will affect the expected signal shape, the relative utility of different frequency bands, the telescope pointing positions on the lunar disk, and most fundamentally, the chances of detecting the known UHE cosmic ray and any prospective UHE neutrino flux. Near-future radio-telescopes such as FAST and the SKA promise to be able to detect the flux of cosmic rays, and it is critical that surface roughness be treated appropriately in simulations. of the lunar Askaryan technique.
In this contribution, a facet model for lunar surface roughness is combined with a method to propagate coherent radio pulses through boundaries to estimate the full effects of lunar surface roughness on neutrino-detection probabilities. The method is able to produce pulses from parameterised particle cascades beneath the lunar surface as would be viewed by an observer at Earth, including all polarisation and coherency effects. Results from this calculation are presented for both characteristic cosmic ray and neutrino cascades, and estimates of the effects mentioned above - particularly signal shape, frequency-dependence, and sensitivity - are presented.
\end{abstract}

\maketitle

\section{Introduction}

All current \citep{2010PhRvD..81d2003J,2012ASTRA...8...29B} and future \citep{2014arXiv1408.6069B,2013JPhCS.409a2096R,2012NIMPA.662S..26M} experiments aiming to use the upper lunar surface layers as a medium for ultra-high-energy (UHE) particle detection via the Askaryan effect involve observing 
the Moon from large distances, either via satellite, or from the Earth itself. While Askaryan pulses are characterised by their broadband coherence, linear polarisation, and very short duration, any pulses of radiation which might be detected by these experiments must first pass through the rough lunar surface. Current models of the transmission process, as used in simulations to estimate experimental sensitivity \citep{2012ASTRA...8...29B,2003astro.ph.10295B,2009APh....30..318J,2009ApJ...706.1556G,2010APh....34..293J}, only deal with the large-scale effects of surface roughness in refracting a pulse. It has also recently been shown that the lunar Askaryan technique will be sensitive to cosmic-ray interactions \citep{2011PhRvE..84e6602J,2010PhRvD..82j3014T}, and consequently these particles have become the target of proposed experiments with the SKA \citep{2014arXiv1408.6069B} --- yet previous estimates of lunar surface roughness effects have only modelled neutrino interactions \citep{2012NIMPA.662S..12J,2009APh....30..318J}. The goal of this contribution is to address this issue.

A toy model of a $100$~EeV hadronic cascade is developed in Sec.\ \ref{SecCascade}, and its radio-emission in an infinite medium compared to that from theoretical parameterisations. This source model is then used as input to the simulation chain described by \citet{2012NIMPA.662S..12J}: briefly, radiation from this cascade is calculated according the \citet{1992PhRvD..45..362Z}, using characteristic dielectric properties of the Moon \citep{1975E&PSL..24..394O}. Artificial lunar surfaces are generated according to \citet{1995JGR...10011709S}, with transmission through surface facets based on the methods of \citet{1939PhRv...56...99S}. In Sec.\ \ref{SecTesting}, the application of this code to near-surface cascades is tested against theoretical expectations, while in Sec.\ \ref{SecResults}, some sample Askaryan pulses are produced, and their properties analysed, to determine the feasibility of cosmic-ray detection using the lunar Askaryan technique.

\section{Model of particle cascades and radio emission}
\label{SecCascade}

The only simulations of the radio-emission from particle cascades in lunar regolith have been performed by \citet{2006PhRvD..74b3007A}, using the `ZHS' code \citep{1992PhRvD..45..362Z}. This code simulates the full $4$D behaviour of cascade particles; while the output is only valid in the far-field of the source, correct results can be attained by breaking up the particle-tracks into a sufficiently small number of sub-tracks, up to a very-near-field limit \citep{2013PhRvD..87b3003G}.

While using a full 4D cascade as input to the roughness calculation of \citet{2012NIMPA.662S..12J} is computationally prohibitive, it has also been noted by \citet{2000PhRvD..62f3001A} that, due to the much greater longitudinal than lateral dimensions of a cascade, applying the radiation calculations of \citet{1992PhRvD..45..362Z} to the longitudinal excess-charge distribution in a cascade produces the observed emission everywhere except very near to the Cherenkov angle, where the lateral extent of the cascade becomes important. Such a calculation is known as the $1$D approximation. In this work, such a $1$D approximation is used to model particle cascades in the regolith. Note that while the lateral dimensions should interact only weakly with surface roughness effects, this is yet to be explicitly tested. Of particular importance is the offset of the cascade from the initial interaction point (for cosmic rays, the lunar surface). This should be a marked improvement on previous estimates using a $3$~m `boxcar' profile \citep{2012NIMPA.662S..12J}, although an improved model, based on the more recent results of \citet{2012APh....35..287A}, should be developed (e.g.\ the scaling obviously produces an offset which is too great).

\begin{figure}
\includegraphics[width=0.4\textwidth]{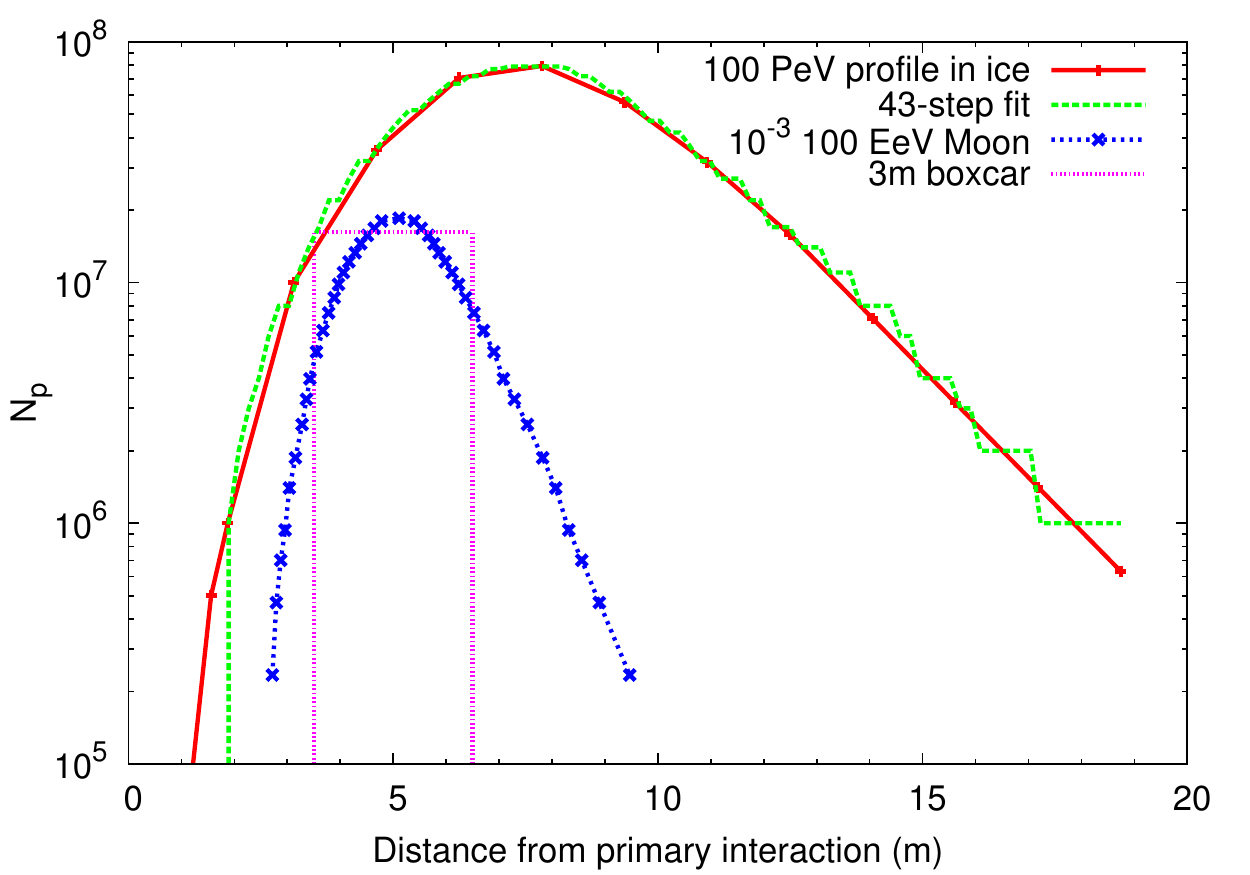} \includegraphics[width=0.4\textwidth]{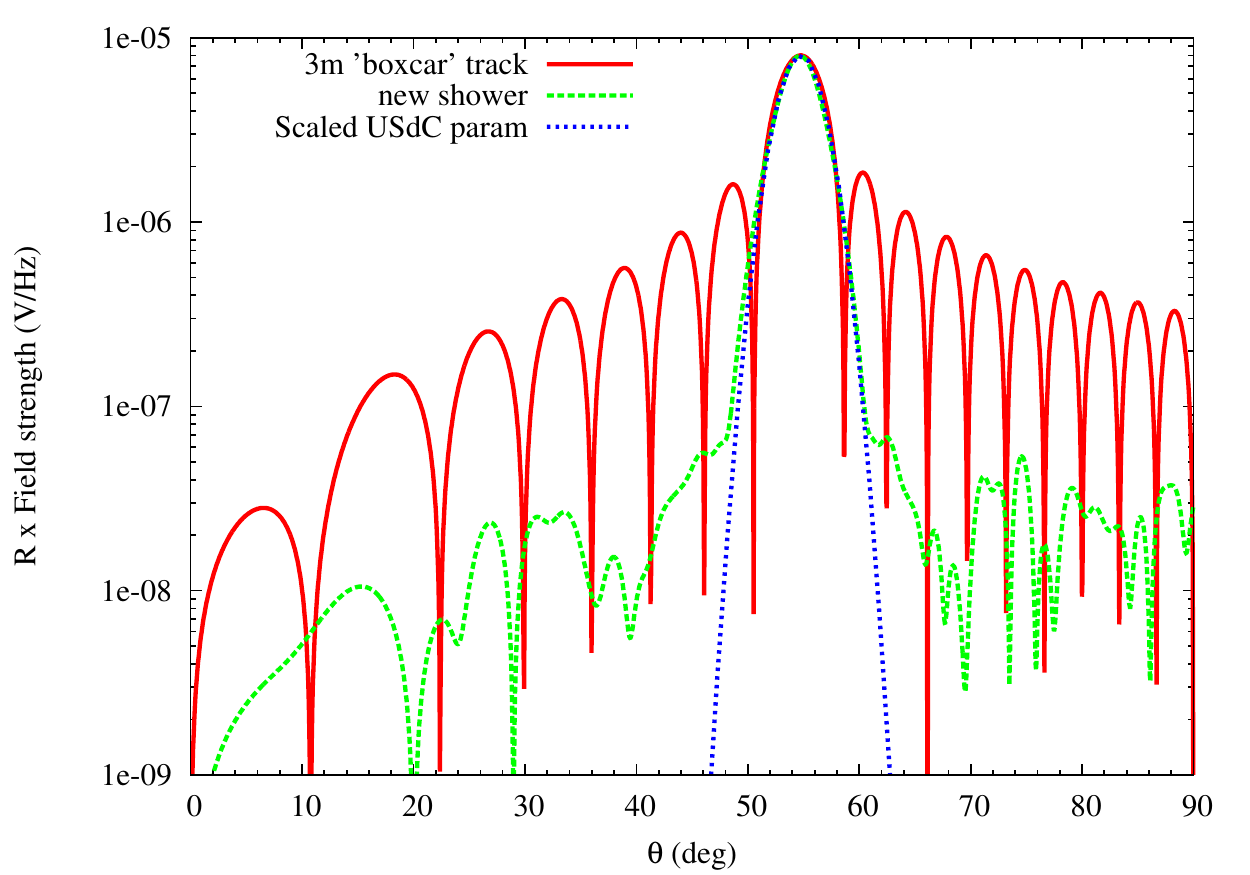} 
\caption{(left) Red: leptonic ($e^- + e^+$) profile of a $100$~PeV cascade in ice \citep{1998PhLB..434..396A}; green: a $43$-step fit to the profile; blue: the excess tracklength ($e^- - e^+$) profile of the scaled $100$ EeV lunar cascade, reduced by $10^{3}$ for clarity; and pink: the $3$~m boxcar profile used in previous simulations. (right) Angular spectrum of radiation produced in regolith by the `lunar cascade' and `boxcar functions', compared to theoretical predictions \citep{1998PhLB..434..396A,2009APh....30..318J}.}\label{FigTrackRad}
\end{figure}

In order to characterise hadronic cascades from the regolith, profiles of the $e^+ + e^-$ longitudinal distribution generated by a full simulations in ice at $100$~PeV \citep{1998PhLB..434..396A} were parameterised with a small number ($43$) of tracks of equivalent charge $Q \propto N(e^+ + e^-)$. In order to scale these results for $e^+ + e^-$ in ice to excess $e^-$ for $100$~EeV cascades in the regolith, both the shower length and charge magnitude were scaled to produce agreement with the parameterisations used in simulations \citep{2009APh....30..318J}. The resulting excess-charge distribution is also given in Fig.\ \ref{FigTrackRad}(left), while in Fig.\ \ref{FigTrackRad} (right), the radiation calculation of \citet{1992PhRvD..45..362Z} is applied to this distribution, with the resulting angular spectra being compared to the parameterisations of \citet{1998PhLB..434..396A} as scaled to regolith according to \citet{2009APh....30..318J}. Note that using a smoother profile reduces the radiation sidelobes, allowing a better study of the degree of accuracy of the calculation.

\section{Testing}
\label{SecTesting}

Before proceeding to full rough-surface calculations, the accuracy of the calculation is tested against the only analytically tractable problem, which is one of no transmission at all. This can readily be modelled in the numerical calculation by setting the regolith absorption length to infinity, and the outgoing `vacuum' refractive index to that of the regolith ($1.73$). The limit of such a test is the finite size of the `transmitting' surface, since in order to calculate the field over all outgoing angles, an infinite surface would be required.

\begin{figure}
\includegraphics[width=0.45\textwidth]{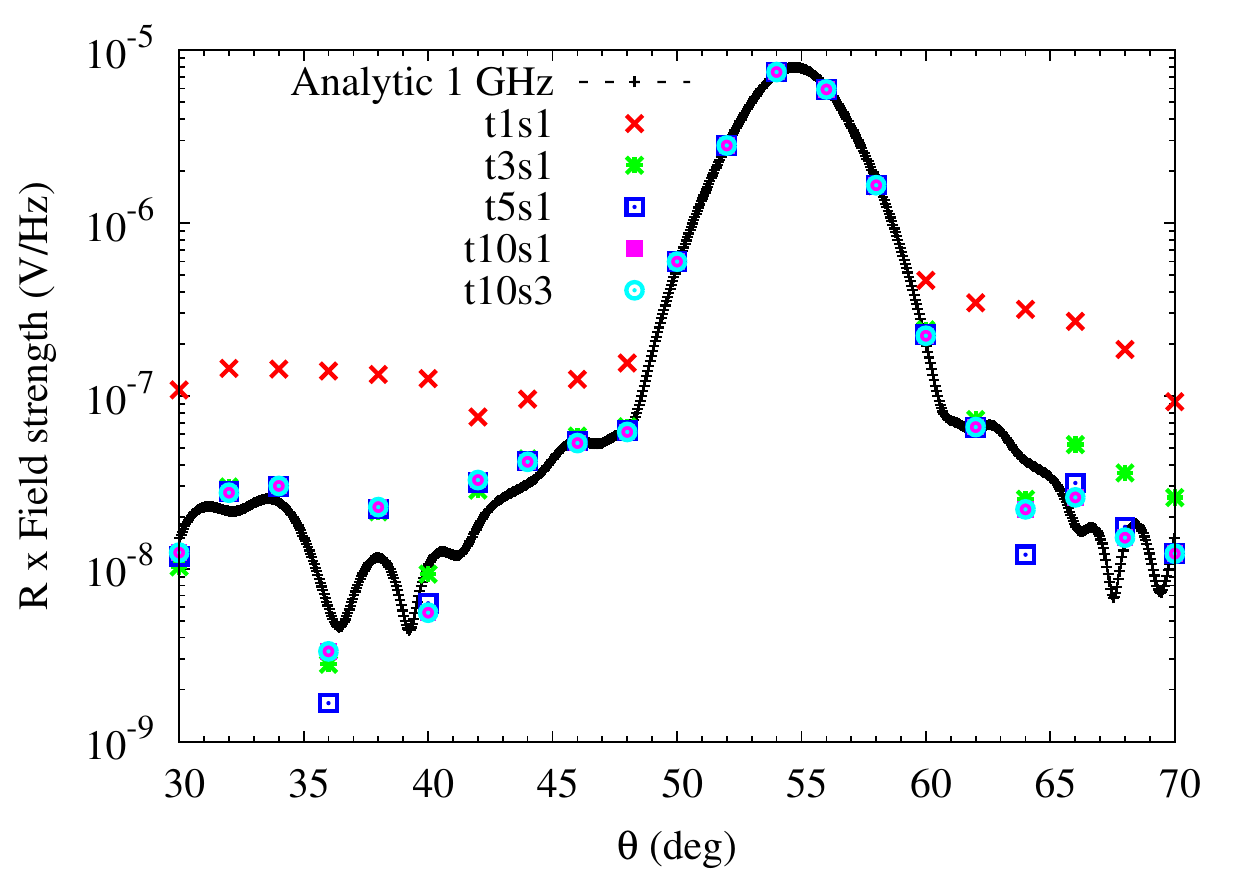} \includegraphics[width=0.45\textwidth]{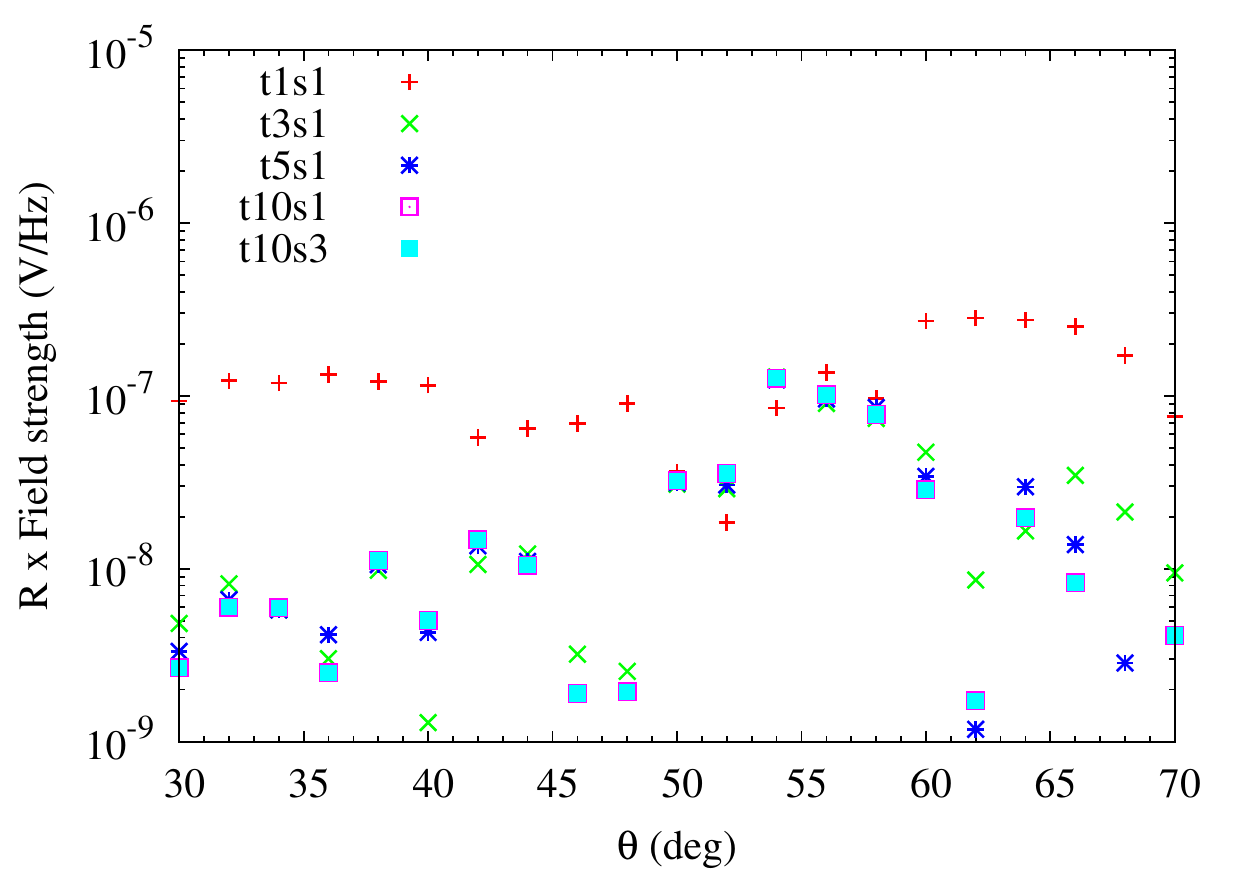}
\caption{(left) Theoretical and numerical calculation of the $1$~GHz emission from the lunar particle tracks in Fig.\ \ref{FigTrackRad} in the case of no refraction; (right) absolute values of the errors with respect to the theory. Label `txsy' indicates that each of the $43$ tracks were divided x times, and each of the $1448^2$ facets were divided y$^2$ times.}\label{FigTest}
\end{figure}

In order to reduce the possible parameter space for testing, only one geometry was chosen: cascades were placed horizontally $1$~m below the surface. The entire surface is made of $1448 \times 1448$ pieces, each $1$~cm$^2$. Since the total surface size ($\sim14\times14$~m$^2$) is not much larger than the dimensions of the track profile of Fig.\ \ref{FigTrackRad}(left), calculations were restricted to outgoing angles $\theta$ between $30^{\circ}$ and $70^{\circ}$. The results were analysed only in the plane of the shower axis and the bulk surface normal at a frequency of $1$~GHz, with results shown in Fig.\ \ref{FigTest}. Each numerical point is the sum of at least $90$ million applications of the track-facet transmission calculations outlined in the Introduction.

From Fig.\ \ref{FigTest}, two results are worth highlighting. Firstly, for this geometry, the track size is the variable most limiting accuracy, with results still not converged after $10$ divisions ($43 \times 10$ track segments), whereas increasing the number of facet divisions produced no change. This is not unexpected, since at $10$ track and $1^2$ facet divisions, the dimensions of each are approximately equal ($\sim 1$~cm). Secondly, it appears that the absolute accuracy stays approximately constant over three orders of magnitude in theoretical expectation. This is the characteristic accuracy which might be expected for the result presented in Sec.\ \ref{SecResults}, although further numerical studies should be performed in order to improve this.

A final note worth mentioning: \citet{2013PhRvD..87b3003G} note that in the very near field, the methods of \citet{1992PhRvD..45..362Z} break down even for infinitely small track sub-divisions. In such a case, a more detailed calculation might be required of the fields incident upon the surface --- this is left to a future work.

\section{Results}
\label{SecResults}

To estimate the radiation from cosmic-ray cascades in the Moon, the zero-point of the cascade was placed at a random point on an artificial surface, with the cascade pointing downwards at a range of angles. Here, results for only one surface, and angles of $\theta_{\rm CR}=10^{\circ}$ and $30^{\circ}$, are shown. This placed the excess charge maximum at an average depth of $85$~cm and $2.5$~m respectively.

\begin{figure}
\includegraphics[width=0.45\textwidth]{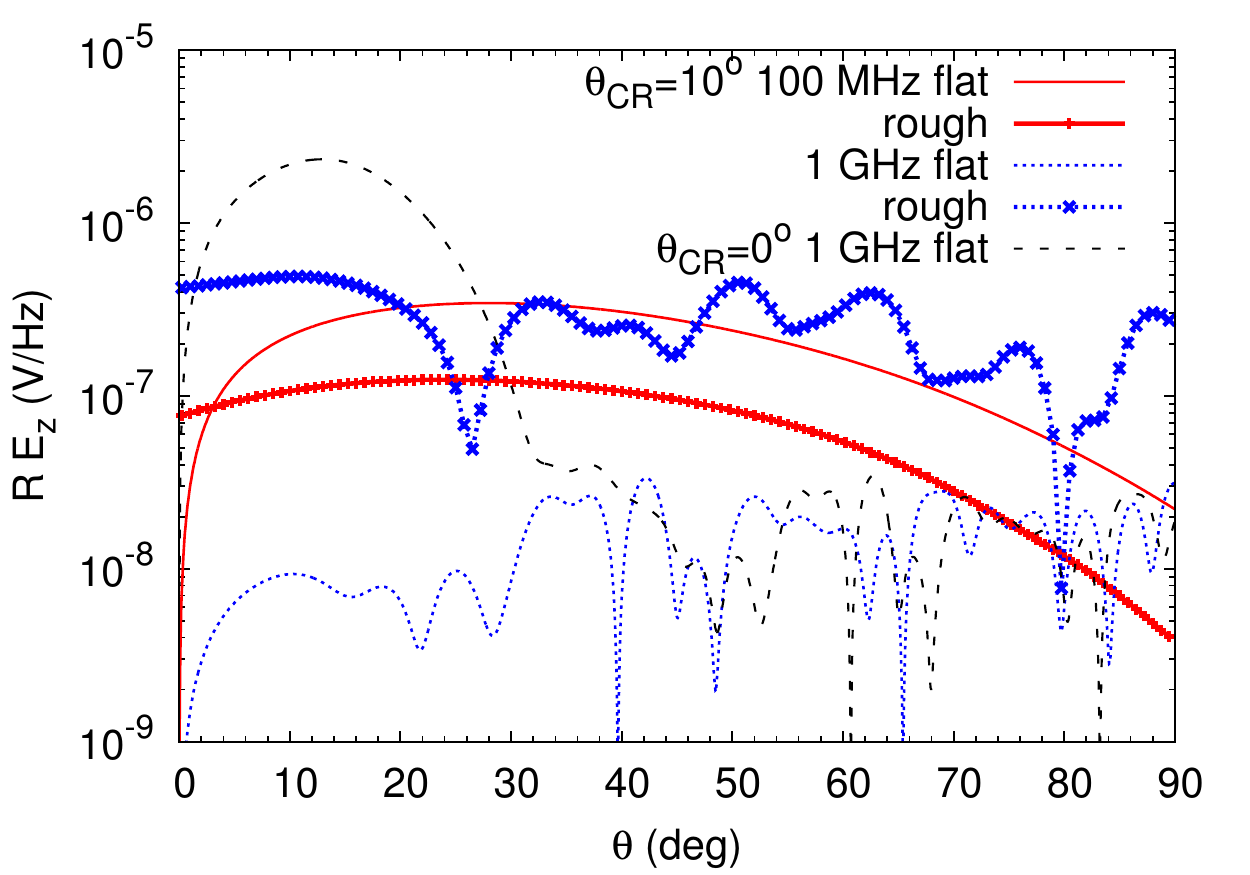} \includegraphics[width=0.45\textwidth]{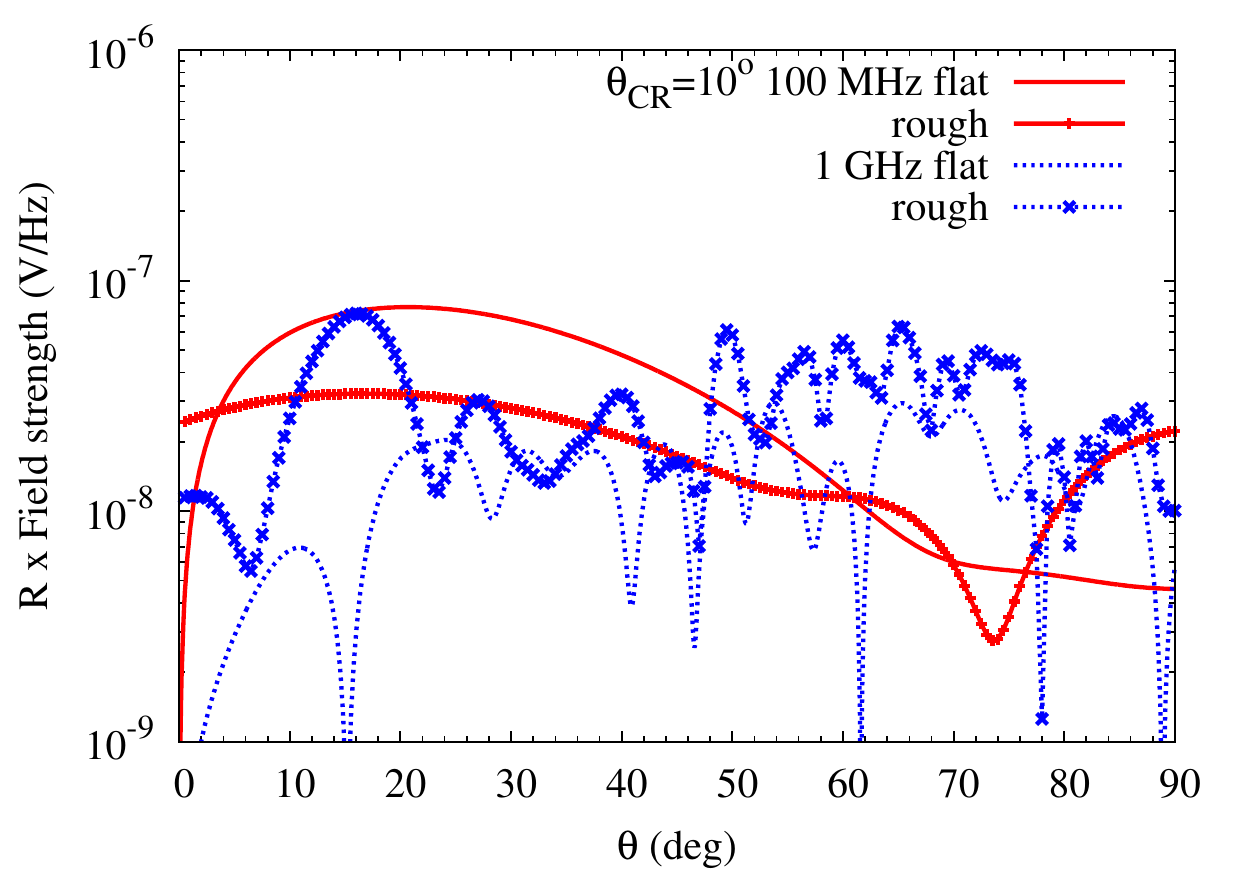}
\caption{Examples of numerical `rough' and theoretical `flat' calculations of transmitted field strength in the vertical polarisation, $E_z$, at $1$~GHz and $100$~MHz for a single sample surface $100$~MHz, for a $100$~EeV cosmic ray incident at $\theta_{\rm CR}=10^{\circ}$ (left) and $30^{\circ}$ (right). Also shown for reference is $\theta_{\rm CR}=0^{\circ}$ at $1$~GHz. }\label{FigResults}
\end{figure}

Results for the vertically polarised electric field strength $E_z$ at $100$~MHz and $1$~GHz for the numerical rough-surface calculation are compared to their analytic equivalents in the case of a flat surface in Fig.\ \ref{FigResults}. Note that the `rough' case also includes large-scale effects which are not (but could be) included in the analytic calculation. Instead, the case for $\theta_{\rm CR}=0^{\circ}$, which is the most optimistic case possible, is also plotted for reference purposes. 

In the particular case shown,  the transmitted field strength at $100$~MHz is reduced compared to the flat-surface case, while the emission at $1$~GHz, which would normally be totally internally reflected, is much higher. No interference patterns are evident at $100$~MHz due to small-scale roughness, while the interference patterns at $1$~GHz are no more pronounced than those in the flat-surface case. The most important effect seems to be that the decoherence effects which reduce the radiated intensity over the longitudinal cascade dimension are `smeared out' by surface roughness, allowing significant high-frequency emission at greater angles from the surface. Compared to the analytic flat-surface case with grazing ($0_{\rm CR}^{\circ}$) incidence, it appears as if the excess power near $\theta=0^{\circ}$ (i.e.\ the refracted Cherenkov angle) is scattered to larger outgoing angles.

This should not be taken to be indicative of the usual behaviour, but simply an example of small-scale surface roughness effects. What is important however is that significant emission escapes the surface over the entire frequency range, and that the GHz emission can in fact be stronger than at $100$~MHz, where analytic estimates would suggest much weaker emission.

\section{Conclusions}

It is now possible to model the effects of surface roughness on lunar cosmic ray detection, although some small uncertainty remains concerning the applicability of the ZHS formula for such near-surface problems. The optimum numerical settings are not yet fully worked out, although track pieces of comparable size to the surface facets appear to produce optimal results. With both at $1$~cm, an accuracy of better than $1$\% of the peak amplitude appears achievable, and improvements are expected in the future. The preliminary results presented indicate that small-scale surface roughness produces more significant emission at larger outgoing angles than in the case from flat-surface calculations, particularly at higher frequencies, which has implications both for the optimum choice of frequency range, and for the beam-pointing positions of experiments. Further, more comprehensive (and computationally intensive) work is needed to determine the effect on the effective experimental aperture to ultra-high-energy cosmic rays.

\bibliography{cwjames_roughness}{}
\bibliographystyle{aipproc}

\end{document}